\pdfoutput=1
\documentclass[11pt]{article}

\usepackage[preprint]{coling}

\usepackage{times}
\usepackage{latexsym}
\usepackage{amsmath}
\usepackage[utf8]{inputenc}
\usepackage{tabularx}

\usepackage[T1]{fontenc}
\usepackage[utf8]{inputenc}

\usepackage{microtype}

\usepackage{inconsolata}

\usepackage{graphicx}

\title{A Framework for Ranking Content Providers Using Prompt Engineering and Self-Attention Network}

\author{Gosuddin Kamaruddin Siddiqi \\
  Microsoft  \\
  \texttt{gsiddiqi@microsoft.com} \\\AND
  Deven Santosh Shah \\
  Microsoft  \\
  \texttt{devenshah@microsoft.com} \\\And
  Radhika Bansal \\
  Microsoft  \\
  \texttt{rabansal@microsoft.com} \\\And
  Askar Kamalov \\
  Microsoft  \\
  \texttt{askarkamalov@microsoft.com}
  }

\begin{document}
\maketitle
\begin{abstract}
This paper addresses the problem of ranking Content Providers for Content Recommendation System. Content Providers are the sources of news and other types of content, such as lifestyle, travel, gardening. We propose a framework that leverages explicit user feedback, such as clicks and reactions, and content-based features, such as writing style and frequency of publishing, to rank Content Providers for a given topic. We also use language models to engineer prompts that help us create a ground truth dataset for the previous unsupervised ranking problem. Using this ground truth, we expand with a self-attention based network to train on Learning to Rank ListWise task. We evaluate our framework using online experiments and show that it can improve the quality, credibility, and diversity of the content recommended to users. 
\end{abstract}

\section{Introduction}

News aggregators have an impact on society and individual news provider \citep{athey2021impact}. Recent studies showcase that news aggregation exhibits a symbiotic relationship between the aggregator and news providers \citep{lee2015rise} \citep{sutanto2021future}. These news aggregators have evolved and are not limited to just news reporting type of content. Microsoft News and Feeds\footnote{Microsoft News and Feeds Product: https://ntp.msn.com/edge/ntp?locale=en-US} is one such evolved product that showcases news and content beyond news such as lifestyle, travel, gardening, food and recipes. Content Aggregators faces repercussions as they experience content reporting or coverage about the same events and happenings from multiple content providers \citep{taurino2022open}. In the age of increased digital accessibility and ease of spread of misinformation, content aggregators face an uphill battle to keep their product updated, informative and at the same time reliable and trustworthy \citep{adema2010service}. This paper focuses on utilizing user feedback such as clicks, explicit reactions such as like and dislikes and other properties of Content such as Writing and Information Quality, and sentiment portrayed in the Content. This paper demonstrates a novel solution to the problem of ranking publishers through information retrieval and text mining and solving for bias and promoting fairness to a certain extent. The paper also demonstrates usage of Prompt Engineering to achieve scale over an unsupervised ranking problem. The outcome of this paper is a framework that can help organize and rank content for various use cases such as Content Recommendation System. 

\section{Literature Review}

To differentiate between different content providers, Media Rank, \citep{ye2019mediarankcomputationalrankingonline} developed a framework that helps rank Online News Sources. Media Rank focuses on four dimensions, Popularity, Reporting Bias and Breadth, Peer Reputation and Financial bottom-line pressure of ads and social bots. Media Rank focuses only on the features curated from the content itself and does not consider the explicit user feedback. Also, Media Rank crawls over the Internet to curate the Online News Sources that they would want to rank and differentiate. Media Rank provides a flat ranking of all the Content Providers which can be further differentiated by identifying the best topic for a given provider for example ESPN will be ranked against New York Times but there needs to be an explicit differentiation or identification of topic that New York Times or ESPN publishes on. This is an additional task on the adopters of Media Rank rankings. Moreover they do not cover all the possible topics or do not provide an exhaustive list of topics for which these rankings are good for. Another source of content provider ranking is News Guard\footnote{NewsGuard: https://www.newsguardtech.com/} which claims to provide rankings for a set of topics but the technique is not Open Source and serves as a black box for their adopters. News Guard ranking is based on the Reliability Rating generated by curating over 30 data points for each News Source \citep{newsguard2021misinformation,newsguard2023aitracking,newsguard2021researchers}. Recent report from Reuters \citep{reuters2022digital} provides a comprehensive overview of consumption of News through different sources of media including popular social networks. The report entails well organized information about sorting News Sources on different source of media offline/online by a Brand Trust Score.

In order to resolve the challenge, several solutions are proposed from Learning to Rank perspective that range from methods as simple as curating signals to Deep Neural Networks. \citealp{pobrotyn2021contextawarelearningrankselfattention} proposes a context-aware self attention network at inference time for user-item interaction, to account for other ranked candidates. \citealp{li2022learning} explored the idea of relationship between ranking features using multi-head self-attention mechanism with Conditional Generative Adversarial Nets (CGAN) to provide a solution for this problem.

\section{Problem Structuring}
Content Aggregators faces an arduous challenge to recommend or surface content from best Content Providers \citep{lee2015rise}. Since Content Providers are contractually bound, apart from what is available as Open Source, not many features, information or knowledge is available to create a framework to solve the problem \citep{taurino2022open}. As new Content Providers are developed, or onboard on to Content Aggregators, it acts as cold start problem. Unless there is any information shared explicitly by Content Providers, Content Aggregators rely on internal data extracted from Content Understanding and/or Explicit User Feedback. The scope of Content Aggregators may or may not be limited to a set of languages or regions. This is an additional complexity and dimension to consider while choosing the best Content Providers \citep{ye2019mediarankcomputationalrankingonline}. This culminates to an unsupervised ranking problem, where the true rankings are unknown, even in context of a specific Content Aggregator.

Microsoft News and Feeds Product has around 50+ languages and regions, and 5200+ unique Content Providers publishing on 4300+ topics. Each Content Provider does not necessarily map to publishing content about one topic or language and region alone, which increases the cardinality. For example, BBC, publishes content on multiple broader topics such as World Politics, Breaking News events, on technology, showbiz world, sports, and business\footnote{BBC: https://www.bbc.com/}. Harper Bazar, is focused on a set of topics, such as Fashion, Beauty, Celebrities and Culture\footnote{Harper Bazar: https://www.harpersbazaar.com/}. ESPN on the other hand is solely emphasizes on Sports news\footnote{ESPN: https://www.espn.com/}. Media Rank \citep{ye2019mediarankcomputationalrankingonline} has a fully automated algorithmic ranking 50,000 Online News Provider, which does not completely overlap with our Content Recommendation System. For example, a well-renowned News Provider, The Times of India \citep{reuters2022digital}, is not available in Media Rank rankings. This suggests that even though Open Source and multiple sources of truth available, ranking Content Providers, is an \textbf{unsupervised ranking problem}, Content Aggregator specific problem to solve. 

\section{Formulating solution}
For each Content in the Content Recommendation System, Content Provider Ranking is one of the many signals used for ranking the Content to create a homogeneous looking Content Feed. Each Content is published by a Content Provider for a given language and region. Content Catalog is semantically clustered based on theme of the article, referred to as “Topic” of the Content. There can be multiple topics that can be inferred for a content to better describe it. For example, a content with combination of topics “gardening”, “summer”, “sunlight”, will be different from a content with combination of topics “gardening”, “greenhouse”, “global warming”. In the current recommendable catalog there are 4,300+ topics on a daily average. Since a content can have multiple topics, individual topics have a confidence score, implying topics with high score has high semantic similarity between content and topic. 

Our solution generate ranking for Content Providers in all observable Topics, language and regions. In practice for recommending, at content level, we take the best 3 topics sorted by confidence score and curate the generated ranking for the combination of Topic, Content Provider and Language and Region, average them to get a final ranking for the content. In our Content Recommendation System, at content level, we identify Publisher Topic Ranking as property and feature of Content. As discussed earlier, a Content (\(C\)) can identify with multiple themes or topics (\(T_{1}, T_{2}, ... , T_{n} \)) and will have one Content Provider (\(P_{C}\)). Since Publisher Topic Ranking is derived for Topic - Provider pair, for each Content, we take the best 3 topics (\(T_{1}, T_{2}, T_{3} \)) and aggregate the signal (\(avg(P\text{x}T_{T_1 P_{C}},P\text{x}T_{T_2 P_{C}},P\text{x}T_{T_3 P_{C}})\)) as an average to infer a final ranking for that Content (\(P\text{x}T_{C}\)). This signal is used in conjunction with existing signals as a multiplicative function in the recommendation system. 
\[
f(x) = f_{1}(y_{1}) * f_{2}(y_{2}) * P\text{x}T_{C} * ... *f_{n}(y_{n})
\]

In the following subsections we detail the evolution of Content Provider ranking through different signals and methods. A combination of Content Based and Engagement Based features are used in determining the ranking. Hence the ranking is dynamic in nature and expected to be updated based on the Content Based Features such as quality of the coverage of the content by the Content Provider, and in turn engagement response from the users.

\subsection{Weak Supervised Approach}

Weak Supervised Approach is more statistical driven given the nature of the problem is that of unsupervised ranking. The cost of labeling all the possible topics is very expensive. We take into account a set of familiar and high engaging topics for English language and USA market that is more familiar with the authors. Considering this, the first technique is more of exploratory framework. A score is generated as a Linear Combination of all majorly user explicit feedback and content based features mentioned in the Appendix B. When sorted by the score the best rank is rank 1 has the highest score while the worst rank has a score close to 0. In order to keep the freshness of the rankings, we continuously update our features on a daily cadence. To avoid noise from daily updates, we take a weighted average of each day for a given feature, with more weights to recent most day's value. With statistical observations, we come up with a set of weights assigned to each features that can help replicate a ranking similar to Copilot prompt\footnote{Copilot: https://www.bing.com/chat?form=NTPCHB} mentioned in Appendix A. To validate the weights we create a small prompt over Copilot to identify if the rankings generated from these weights are aligned. The idea of using external prompt represents a Knowledge proxy through the topic and the Content Providers in the current given context of News and Feed content recommendation system. The response from Copilot prompt is available in table \ref{tab:content_providers} Appendix A, and the corresponding weights adjusted generated ranking is available in table \ref{tab:wts_generated_ranking} Appendix B.

\subsubsection{Shortcomings}

The usage of external Copilot prompt presents another difficulty of coming up with concrete metrics to measure the effectiveness of the prompt and the weights because the Copilot prompt will consider all possible Content Providers for ranking not limited to our Content Recommendation system. Semantic meaning of topics for Copilot prompt and the internal data will correlate but not necessarily be the exact same, hence we can observe some deviation in judging the rankings. For example, based on generated rankings, \textit{E! News\footnote{E! News: https://www.eonline.com/news}} is a top Content Provider for Sports and Health. The intuition is, Content Provider is likely publishing content about a celebrity linked with Sports and covering their health in recent time.

\subsection{Prompt Engineering and Self-Attention Network}
Prompt Engineering and Self-Attention Network is our ideal solution framework proposed as an outcome of this paper. In this approach, we strive to make a ground truth data set using human judgments and expand with the help of detailed prompt engineering using GPT-4o. As for the first step we asked our Subject Matter Experts (SME) to select all the Content Providers that they consider are best for a given topic. The candidates for this task consist of both positive labels and the randomly sampled negative labels to avoid any biases while selecting top Content Providers. For positive samples, we take the top 30 Content Providers from the previous iteration and for negative samples, we randomly select 10 to 15 Content Providers. We ask multiple annotators, here SMEs, to select top Content Providers. Alongside, we create a comprehensive prompt with exhaustive guidelines and suggestions to identify and rank the best Content Provider and add or remove Content Provider not in the input list. These guidelines are created keeping in view point to create long term value for the users of our Content Recommendation System, that promotes, fair, high quality and engaging content for each top Content Provider. We reach consensus (2 out 3 agreement) from multiple annotators, when majority of them agree that a give candidate Content Provider is in top list for a given topic. We continuously iterate our prompt engineering to match results from human judgement. Upon high agreement of Precision/Recall of 95/95 between prompt and human judgement, we finalize the prompt and scale it to all topics, languages and regions.

Using this prompt, we achieve scalability as opposed to Weak Supervised Approach where we limited the truth set to few topics and English Language and USA region. Since the truth set now exists, we can help reduce the noise from previous approach, and stabilize the rankings. Counter to this solution from prompt results, we face a cold start problem for our new onboard providers. we observe that newly onboard Content Providers will suffer since we lack the labels and hence there is a need to rerank the Content Providers for all the topics that it is publishing content for. To solve this issue, we train machine learning models, help predict the relevance of the new and old Content Providers for a given topic.

We use a common set of features in the following two Machine Learning models:
\begin{itemize}
  \item GPT Ada V2 embedding for Content Provider's Brand Mission Statement
  \item GPT Ada V2 embedding for Topic Definition
  \item Set of Content Based and Engagement Based cross sectional features for a given combination of Content Provider, Topic and Language and Region, collectively called as Numeric Features.
\end{itemize}

\subsubsection{LightGBM Model - Pairwise Ranking}
\begin{figure}[t]
  \includegraphics[width=\linewidth]{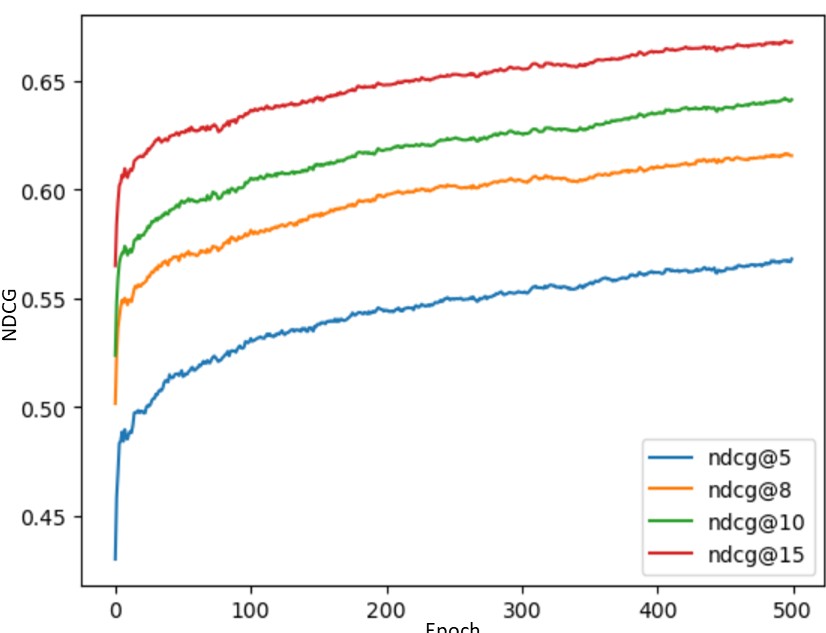}\hfill
  \caption{NDCG gains when training with only positive samples.}
  \label{fig:lgbmonlypositive}
\end{figure}

\begin{figure}[t]
  \includegraphics[width=\linewidth]{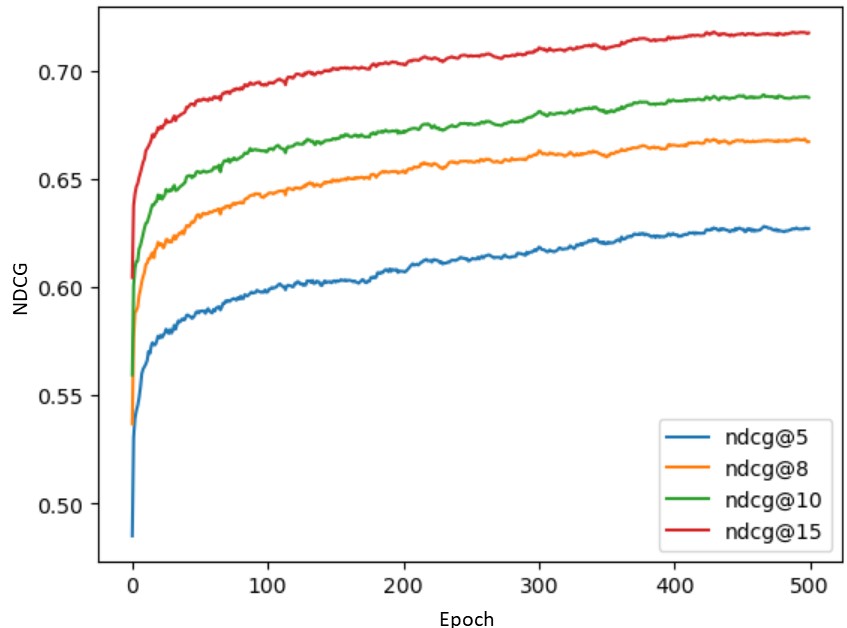}\hfill
  \caption{NDCG gains when training with including negative samples with relevance label as 0.}
  \label{fig:lgbmwithnegative}
\end{figure}

We choose LightGBM Ranker - Pairwise Model, for the effectiveness, efficiency and scaling properties \citep{ke2017lightgbm}. It also possess the ability to treat missing values as a separate category very swiftly. Relevance Labels provided to the model as part of Learning To Rank task, is an inverted ranking of Content Providers for a given topic. Rank 1 Content Provider is given highest relevance, let's say 25, and with increasing ranks, the relevance drops. With LightGBM we experimented with two different settings, where we used only ranked labels and the other we use a set of randomly sample non-ranked negative sample. These negative samples were given relevance label as 0. We observed that Normalized Discounted Cumulative Gain (NDCG), was higher when using the negative samples \ref{fig:lgbmwithnegative} in the training data as compared to model with only positive samples \ref{fig:lgbmonlypositive}.

We observe that even with higher number of epochs and parameter tuning, we are able to achieve higher performance based on NDCG metrics.

\subsubsection{Self-Attention Neural Network - ListWise Ranking}
\begin{figure}[t]
  \includegraphics[width=\linewidth]{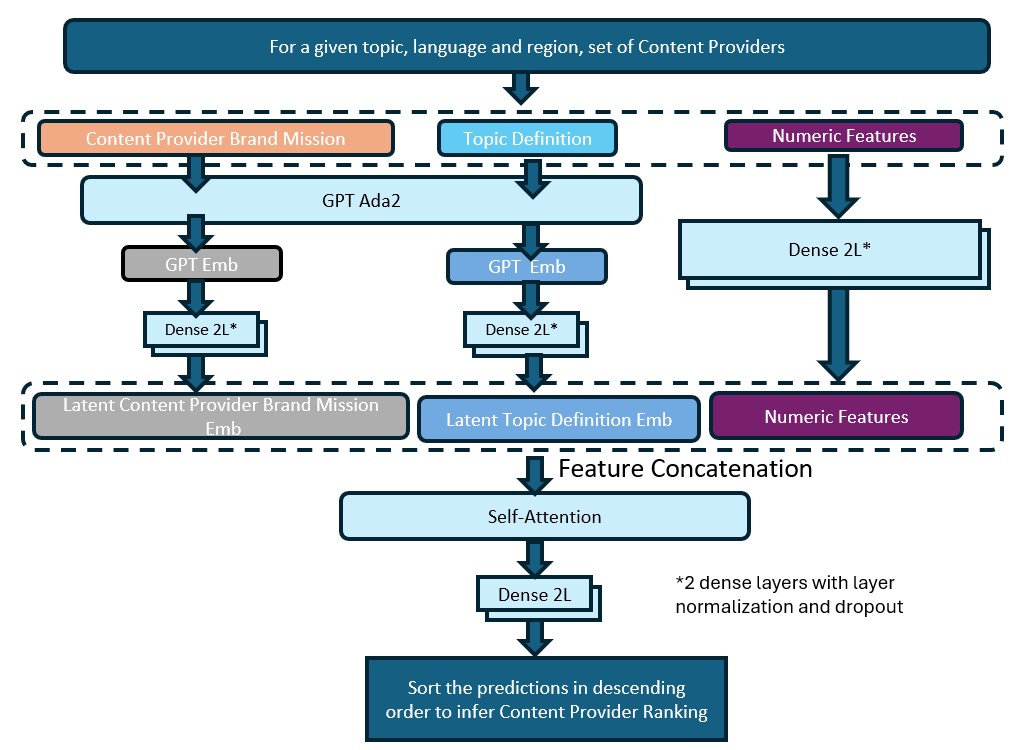}\hfill
  \caption{Self-Attention Neural Network Architecture with ListWise Ranking}
  \label{fig:dnn_listiwse_arch}
\end{figure}

We further propose to Self-Attention \citep{vaswani2017attention} based Neural Network, to extract more latent knowledge from the ADA v2 embedding. We also experiment with our task by asking the network to learn the ListWise Ranking \citep{cao2007learning} task instead. We borrow the learning of using negative samples to supplicate our learning by improving NDCG metric. The architecture of our neural network \ref{fig:dnn_listiwse_arch} comprises of dense layer for each set of features mentioned above, followed by a Self-Attention layer over the concatenated set of features. Upon this we add a set of Dense Layers to induce a list of relevance scores.

\begin{figure}[t]
  \includegraphics[width=\linewidth]{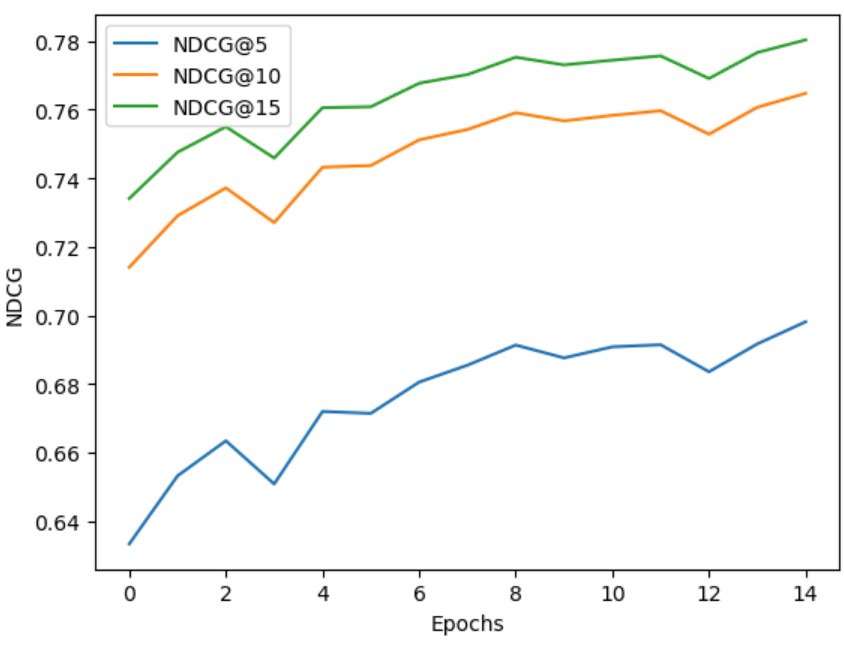}\hfill
  \caption{Self-Attention Neural Network Architecture with ListWise Ranking}
  \label{fig:dnn_ndcg}
\end{figure}

For training our network, we created batch sizes of 64 set of topics, each containing a set of 30 Content Providers which is our slate length \citep{cao2007learning} in ListWise Ranking task. We train up to 15 epochs for our experimental purpose and observe better gains that LightGBM \ref{fig:dnn_ndcg}. 

To summarize our Prompt Engineering and Self-Attention Network approach, we first curate SME judgements over a set of topics, and concurrently engineer a GPT 4o prompt to scale while matching the Precision and Recall of human judgements. To address cold start problem for newly onboard Content Providers, we develop a Self-Attention based Neural Network to infer relevance scores.

\subsection{Experiment Setup}
The effectiveness of new signals in our Content Recommendation System, is determined through A/B testing \citep{xu2015abtesting}. The unit of division is at user level. A user can be either Treatment Group or Control Group at the time of Experimentation. We conduct the experiment across all our Languages and Regions at the same time and evaluate over a period of 7 days. Overall goal of the experimentation is to determine if the new signal can create long-term value for users and business. There are several metrics to measure this, some are more direct such as converting Cold Users as our customers, increasing their interactions and explicit feedback, such as likes and dislikes over recommended content, some are indirect such as stickiness of the users over a period of time. Other essential metrics helps determine overall attractiveness of the feed, such as relevant content, bias and fairness of opinion, positive and uplifting sentiment content, and recommending reliable and trustworthy content.

The setup of experiments are designed to be incremental in nature (see Table \ref{tab:expsetup}). For our Experiment 1 (E1), in the treatment we introduce the signal in multiplicative function, while control is the production system. For Approach 1 - Weak Supervised Approach, the control is without any publisher topic signal. On observing exceptional gains, we add the Experiment 1 treatment in production. Given the incremental nature, treatment for Experiment 2 (E2) replaces signal from E1 and control is now the new production with signal in E1. The proposed Experiment 3 (E3), treatment will augment signal specifically designed for Local Publishers. Control for E3, will be the treatment from E2.
\begin{table}
  \centering
  \begin{tabular}{lcc}
    \hline
    \textbf{Experiment} & \textbf{Control} & \textbf{Treatment} \\
    \hline
    E1   &Production &Approach 1 signal  \\
    E2    &E1 Treatment &Approach 2 signal     \\
   \hline
  \end{tabular}
  \caption{Experiment Setup in incremental iterations}
  \label{tab:expsetup}
\end{table}

\subsection{Experiment Results}

\subsubsection{Experiment 1 (E1)}
Key Results from Experiment 1 (E1) highlights a successful attempt at improving the overall quality and increased positive user engagement from users.
\begin{itemize}
\item Impressions of Content Providers aligned with Brand Mission statement increased by 4.5\% and decreased for not aligned Content Providers by 6.6\% 
\item Daily Active Users on Video Content Type increased by 1.19\% 
\item Impressions on Business critical Content Providers increased by 0.62\% and popular brand increased by 0.28\% 
\item Impression of Bad Quality Content decreased by 0.68\% 
\item Impression of topic Politics decreased by 0.39\% 
\item Impression of Content with predominantly Negative sentiment decreased by 0.21\% 
\end{itemize}

\begin{figure}[t]
  \includegraphics[width=0.48\linewidth]{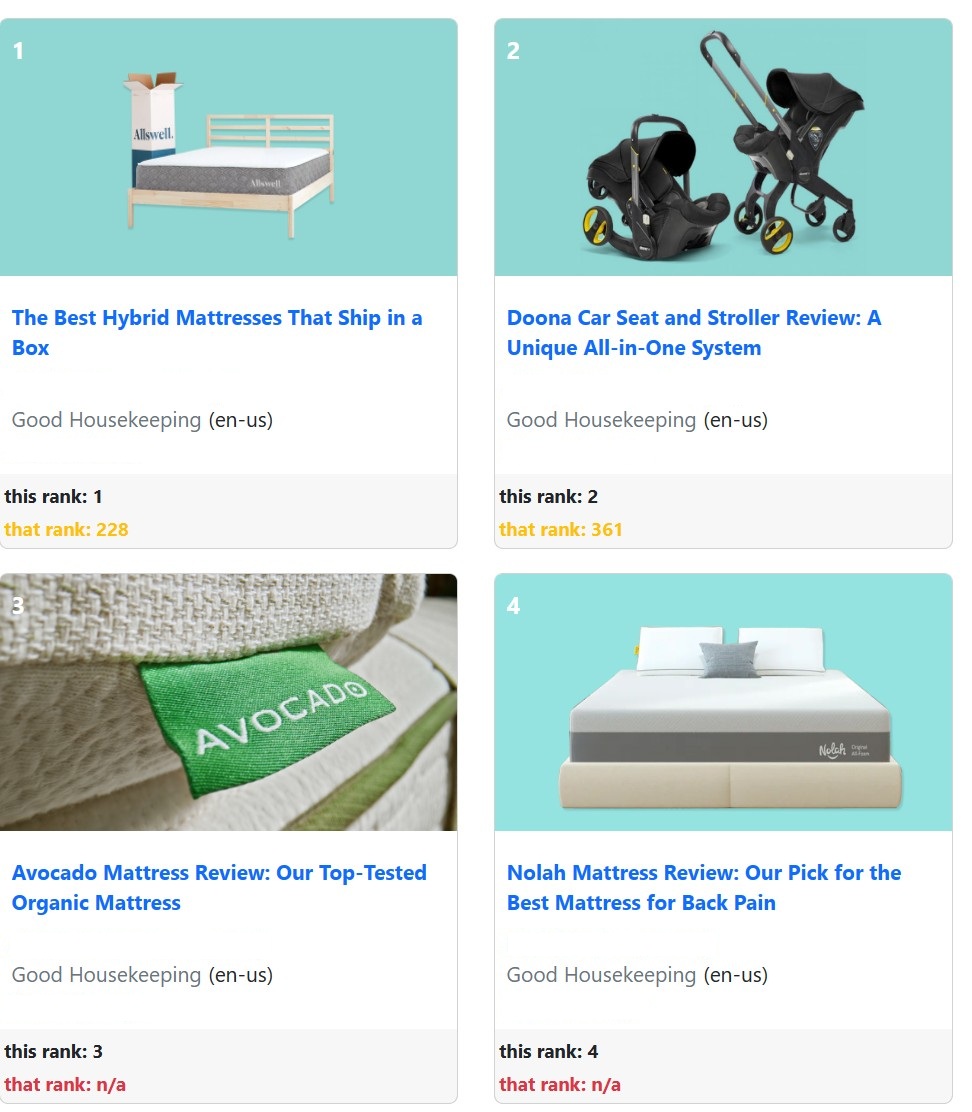}\hfill
  \includegraphics[width=0.48\linewidth]{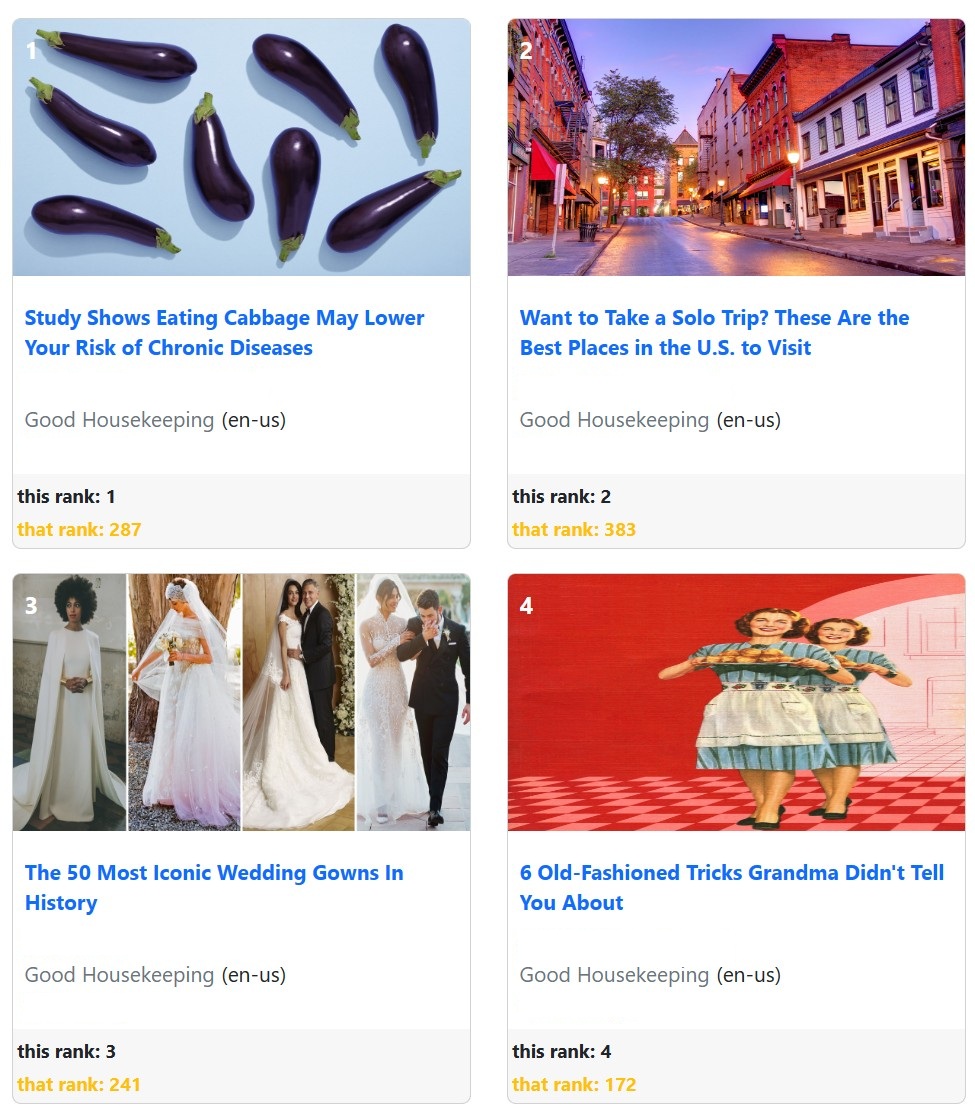}\hfill
  \caption{Side-by-Side candidate generation for a content provider - \textit{Good Housekeeping}. Left half (Treatment) of image showcases better alignment to \textit{Good Housekeeping} brand statement - Good Housekeeping is your destination for everything from recipes to product reviews to home decor inspiration; than the right half (Control). }
  \label{fig:experiments}
\end{figure}

\subsubsection{Experiment 2 (E2)}
Key Results from Experiment 2 (E2) exhibits substantial improvement with the help of Prompt Engineering and Self-Attention Network model, suggesting further improvement in the ranking.
\begin{itemize}
\item Impressions of Content Providers aligned with Brand Mission statement increased by 7\% and decreased for not aligned Content Providers by 3.6\% 
\item Daily Active Users on certain markets increased by 1.03\% suggesting improvements in ranking for those markets
\item Daily Active Users on Video Content Type increased by 1.3\% on first time interacting user
\item Relevance of Recommendations increased by 0.35\%
\item Daily Active Users on Video Content Type increased by 1.8\% on first time visitor
\item Impressions on Business critical Content Providers increased by 0.62\% and popular brand increased by 0.28\% 
\item Impression of Bad Quality Content decreased by 0.31\% 
\item Monetized Content view increased by 0.31\% for first time visitors
\item Impression of topic Politics decreased by 0.66\% 
\item Impression of Content with predominantly Negative sentiment decreased by 0.39\% 
\end{itemize}

\section{Conclusion}
The paper tackles a non-trivial problem for Content Aggregators that ranks Content Providers by considering various dimensions and aspects. The solution to the initial unsupervised ranking problem evolves from weak supervised approach to a ground truth trained model. It also produces a well tested framework that employs a guidelines intensive engineered prompt which is further strengthened by Self-Attention Network model. Online results showcases strong evidences of increased quality at content level and feed level. Increased feedback from end users attests acceptance of recommendations improved using the new ranking of Content Providers.

\section*{Limitations}
Content Recommendation System containing News have a category of Content intended for audiences from a specific geography. This content is generally referred to as Local Content \citep{shah2023whats}. In our previous approaches, for ranking we do not consider Content Providers that are publishing Local Content as this will be a biased source of truth given the popularity of such Content Providers is less than Content Providers publishing only National Content. For example, a regional Content Provider, The Seattle Times\footnote{The Seattle Times: https://www.seattletimes.com/} is likely to be less popular than a National Content Provider, Fox News\footnote{Fox News: https://www.foxnews.com/} or BBC\footnote{BBC News: https://www.bbc.com/news}. Local News is intended to a set of geographically located audience \citep{shah2023locallifestayinformed}, the viewership and user engagement is limited at a smaller scale as compared to national audience. The distribution of user engagement metrics is bound to be different and can result in poor performance for our Self-Attention Network. To solve this, instead of ranking Local Content Providers for each topic, we can rank Local Content Providers considering their geography. We can aggregate Local Content with content level geography \citep{shah2023locallifestayinformed} to a Content Provider, and sort them for example, at city or county level. For instance, The Seattle Times\footnote{The Seattle Times: https://www.seattletimes.com/} will be a candidate for city Seattle, or county - King County. When in unison with ranking for National Content Providers, the gains are expected to be higher with contributions from Local Content.

\bibliography{pxt}

\appendix

\section{Appendix}
\label{sec:appendix}

A. Copilot based Prompt
The prompt used to reinforce the rankings from Iteration 1.

\begin{quote}
"You are a news feed curator who publishes both news and magazine articles, occasionally including timeless pieces that can be enjoyed at leisure. Each article, whether news or magazine style, revolves around a subject or theme. For instance, a news article like '5 Marines aboard helicopter that crashed outside San Diego confirmed dead' discusses an accident. Conversely, a magazine article such as 'See Inside Delta’s Refreshed Cabins — With Revamped Premium Seating, 10-inch Seatback Screens, and More' is timeless and focuses on Air Travel.

News about recent accidents can be covered by multiple publishers in the journalism industry. Your task is to rank these publishers based on their popularity, trustworthiness among readers, article quality, and expertise on a given topic and its broad readership. For each topic I provide, can you rank the top 30 publishers?"
\end{quote}

\begin{table}[ht]
\scriptsize
\centering
\begin{tabular}{|l|l|l|}
\hline
\textbf{World Politics} & \textbf{Fashion} & \textbf{Sports} \\ \hline
The New York Times & Vogue & ESPN \\ \hline
BBC & ELLE & Sports Illustrated \\ \hline
The Economist & Cosmopolitan & Bleacher Report \\ \hline
The Guardian & Allure & CBS Sports \\ \hline
CNN & W Magazine & NBC Sports \\ \hline
Reuters & GQ & Fox Sports \\ \hline
The Washington Post & Vanity Fair & The Athletic \\ \hline
The Wall Street Journal & L’Officiel USA & Yahoo Sports \\ \hline
Al Jazeera & Grazia USA & Sporting News \\ \hline
\end{tabular}
\caption{Prompt Generated Content Providers by Category}
\label{tab:content_providers}
\end{table}

B. Features and Weights used in iteration 1 are highlighted in table\ref{tab:iteration1_features}. The generated ranking is available in table \ref{tab:wts_generated_ranking}.

\begin{table}[ht]
\scriptsize
\centering
\begin{tabular}{|l|c|}
\hline
\textbf{Features} & \textbf{Weights} \\ \hline
Popularity of Content Provider & 0.3 \\ \hline
Brand Mission of Content Provider, Topic & 0.1 \\ \hline
Eligible Article Count in Last 7 Days & 0.05 \\ \hline
High Quality Document Ratio for the Content Provider, Topic & 0.1 \\ \hline
Content Provider Document Ratio for the Topic & 0.1 \\ \hline
Click Dwell Time of Content Provider, Topic & 0.1 \\ \hline
CTR of Content Provider, Topic & 0.1 \\ \hline
User Feedback of Content Provider, Topic & 0.1 \\ \hline
\end{tabular}
\caption{Comparison and weights of Features for Content + Engagement Based.}
\label{tab:iteration1_features}
\end{table}

\begin{table}[ht]
\centering
\scriptsize
\begin{tabular}{|l|l|l|}
\hline
\textbf{Politics} & \textbf{Fashion} & \textbf{Sports} \\ \hline
FOX News & WWD & The Associated Press \\ \hline
CNN & Harper's Bazaar & The Associated Press – Sports \\ \hline
The Associated Press & Glam & BBC \\ \hline
MSNBC & InStyle & SB Nation \\ \hline
Newsweek & People & ESPN \\ \hline
The Washington Post & GQ & CBS Sports \\ \hline
The New York Times & Marie Claire US & NBC Sports \\ \hline
NBC News & Hollywood Reporter & Yahoo Sports \\ \hline
USA TODAY & Cosmopolitan & AFP \\ \hline
The Telegraph & Glamour & E! News \\ \hline
\end{tabular}
\caption{Generated Ranking of Content Providers by Topic}
\label{tab:wts_generated_ranking}
\end{table}

\end{document}